 \documentclass[final,preprint,5p,times,twocolumn]{elsarticle}

\usepackage{amssymb}

\usepackage{url}

\journal{ }

\begin{document}

\begin{frontmatter}



\title{Decomposition and Prediction of Initial Uniform Bi-directional Water Waves Using an Array of Wave-Rider Buoys }


\author{Takahito Iida}
\ead{iida@naoe.eng.osaka-u.ac.jp}

\address{Department of Naval Architecture and Ocean Engineering, Osaka
University, Suita, Osaka 5650871, Japan}

\begin{abstract}
Prediction of incident waves to wave energy converters (WEC) is essential to maximize the energy absorption by controlling the WEC.
Nevertheless, little work has been done on the deterministic prediction of bi-directional waves whose wave directions of components are 180$^\circ$ opposite.
To decompose and predict such bi-directional waves, an array of three wave-rider buoys are considered.
Buoys on both sides are used for decomposing bi-directional waves into progressive and regressive wave components, and the surface elevation of the middle buoy is predicted by these decomposed waves.
The deterministic wave prediction is based on the impulse response function, and a cosine-filtered impulse response function is proposed to reduce an error due to the truncation of the infinite length of the function.
Predictions of initial uniform bi-directional waves are shown to demonstrate the performance of the impulse response function method to time-series prediction.
Both numerical and experimental comparisons are carried out to validate our proposals.
\end{abstract}


\end{frontmatter}


\makeatletter
 \renewcommand{\@cite}[1]{[#1]}
\makeatother

\mathindent=0pt   

\section{Introduction}
The realization of carbon neutrality is set as the world’s most urgent mission\cite{guterres2020}, and the development of renewable energies has become increasingly important. 
Ocean wave energy is one of the promising energy resources which has high energy intensity compared to other resources.
Several concepts of wave energy converters (WEC) have been proposed so far to maximumly extract wave energy on their target sea conditions (see reviews \cite{lehmann2017,al2018,nguyen2020}).
One successful WEC is a point absorber, which consists of a heaving buoy and fixed reference, and its relative motion is converted by a power take-off (PTO) system\cite{al2018}.
Since the mechanism of the point absorber is relatively simple, its construction, operation, and maintenance are easier and cheaper than those of other concepts.
In addition, energy extraction can be increased using multiple point absorbers, i.e. array or farm (e.g. \cite{zhong2019,murai2021,tay2022}).

For maximizing energy absorption, the real-time control of the WEC may be essential (e.g. \cite{ringwood2014,korde2015}), and thus many control methods are proposed, such as feedforward control\cite{naito1986}, latching control\cite{budal1982,babarit2006}, model predictive control\cite{gieske2007}, and so on\cite{hong2014}. 
These real-time controls are effective {\it if the future excitation force is assumed to be known}\cite{hals2011}.
Many studies are based on this assumption, and the wave exciting force is expressed by the superposition of sinusoidal forces with different frequencies (e.g. \cite{ringwood2014,song2016,liu2018}).
However, a prediction of the wave exciting force (i.e. prediction of incident waves) is indispensable to applying these real-time control methods.
A straightforward way of predicting incident waves is a measurement of waves by any sensor arranged in front of the WEC with distance.
Using measured wave profile and impulse response function of water waves, present/future waves incoming to the WEC are predicted\cite{korde2015,naito1986}.
Otherwise, an artificial neural network engages in this wave prediction\cite{desouky2019}.
These up-stream measurements may be especially effective for an array of point absorbers because waves measured at an absorber are used for predicting waves incoming to other absorbers.
Note that these predictions are mostly formulated under uni-directional wave fields or narrow-band bi-directional/multi-directional wave fields\cite{belmont2014,korde2017}.

The most famous approach to the deterministic water wave prediction is a Fourier series coefficient estimation (e.g.\cite{belmont2014,naaijen2018,al2019,al2020}).
This method is suitable for the prediction of the wave exciting force if the force is represented by the superposition of sinusoidal forces.
However, since the Fourier series expansion assumes an infinite length of sinusoidal waves, it may not be appropriate to predict initial uniform waves or transient waves.
Especially, the prediction accuracy of the wavefront is imperfect due to the Gibbs phenomenon.
On the other hand, the prediction method based on the impulse response function can be applied to such wave cases.
Besides, this method is suitable for controlling the WEC by the state-space representation because this method is formulated by a convolution integral.
The analytical solution of the impulse response function for deep water waves\cite{davis1966,falnes1995} is used for the control of the WEC\cite{naito1986,korde2015}.
Nevertheless, this impulse response function is strongly non-causal, and a prediction error may not be acceptable.
Recently, an analytical solution of the impulse response function for finite-depth water waves is also derived\cite{iida2022}.
Although this impulse response function is still non-causal, such non-causal influence can be exponentially reduced with distance\cite{iida2022}.
Moreover, it is indicated that this method can be used to predict future waves with an arbitrary non-causal error when the distance is sufficiently long.

In this paper, we propose the deterministic wave prediction method for bi-directional water waves for the sake of the control of the WEC.
Bi-directional water waves consist of progressive and regressive waves of which wave directions are 180$^\circ$ opposite.
An array of three wave-rider buoys are considered to decompose and predict waves.
Measured wave data at wave-rider buoys on both sides are used to decompose bi-directional waves into progressive and regressive waves.
These decomposed progressive and regressive waves predict waves of the middle buoy.
For the prediction, the analytical solution of the impulse response function under the finite-depth dispersion relation\cite{iida2022} is applied.
In addition, a new analytical solution using a window function (cosine-type) is combined to eliminate a truncation error.
As the impulse response function method is applicable to the prediction of initial uniform waves or transient waves, unlike the Fourier coefficients estimation, predictions of initial uniform wave trains are demonstrated.
We also propose an arbitrary wave generation method based on the impulse response function to make an initial uniform wave profile.
The proposed decomposition and prediction method is validated by both regular and irregular bi-directional waves.
A tank experiment is also carried out.
It is worth noting that the proposed method can be used for detecting reflection waves in the tank experiment, and it may be useful to provide robust experimental results.

\section{Problem description}
\begin{figure}[h]
\vspace*{0mm}
  \centerline{
\includegraphics[width=1.0\linewidth,clip]{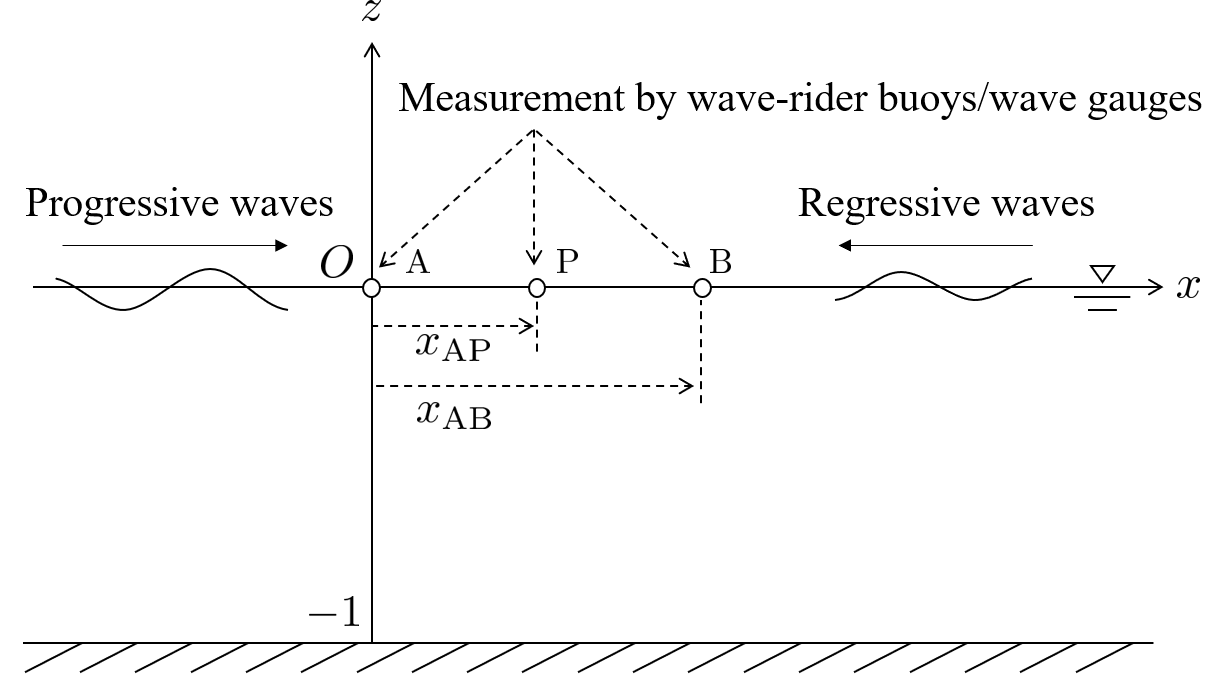}}
\par\vspace*{0mm}
\caption{Schematic view of two-dimensional bi-directional water wave problem. Free surface elevations at three points A, B, and P are measured by wave-rider buoys/wave gauges. Points A and B are used to decompose bi-directional waves into progressive and regressive components. Elevation at point P is predicted using decomposed waves. All values are normalized by water depth, gravitational acceleration, and selected wave amplitude.}
\label{fig1}
\end{figure}

The problem is formulated under a framework of the two-dimensional linear potential flow theory.
Figure \ref{fig1} describes a schematic view of the problem where the origin $O$ is located on the undisturbed free surface and the bottom topography is constant on $z=-1$.
Values are normalized by water depth, gravitational acceleration, and selected wave amplitude.
We consider bi-directional waves which consist of progressive waves (propagating to the positive $x$ direction) and regressive waves (propagating to the negative $x$ direction).
The purposes of the paper are a decomposition of bi-directional waves into progressive and regressive components and a prediction of free surface elevation at the desired point.
To accomplish these purposes, three free surface points A, B, and P are considered.
Point A is located at the origin, and point B is placed on the free surface with the distance $x_{\rm AB}$ from point A.
Besides, point P is arranged between points A and B with the distance $x_{\rm AP}$ from point A.
Points A and B are used to decompose bi-directional waves into progressive and regressive waves.
Using the decomposed progressive and regressive waves, the free surface elevation at point P is predicted.

The main application of our proposal is optimal control of wave energy converters (WEC) in actual sea sites.
To maximize the energy absorption of the WEC, an incident wave prediction is of great importance.
One of the promising approaches is the use of the wave data measured at the up-stream of the WEC\cite{korde2015,naito1986,desouky2019}.
The proposing method expands these harvesting controls to the case of bi-directional waves in which directions of wave components are 180$^\circ$ opposite.
To model the actual sea measurement, wave-rider buoys are considered and arranged at points A, B, and P.
Here, only vertical motion (heave motion) of the buoys is allowed.
Besides, the mass of the buoys is assumed to be sufficiently small; the buoys do not scatter waves and move in contact with the free surface of the water.
Therefore, the wave-rider buoys act like wave gauges.
It is highlighted that the array of the buoys can imitate an array (or farms) of point absorbers.

\section{Prediction based on impulse response function}
The prediction method of bi-directional waves is studied based on the impulse response function method.
Before the bi-directional wave prediction is considered, the prediction of uni-directional waves is reviewed.
Free surface elevation at point P (output: $\xi_{\rm P}(x_{\rm AP},t)$) is predicted by the convolution integral of the elevation at up-stream point A (input: $\xi_{\rm A}(t)$) and the impulse response function $h(x,t)$ as
\begin{eqnarray}
\displaystyle \xi_{\rm P}(x_{\rm AP},t)\approx \int_{0}^{\infty}h(x_{\rm AP},t)\xi_A(t-\tau)d\tau
\label{h:01}
\end{eqnarray}
where integral range $[0,\infty]$ is used assuming the causality of the impulse response function.
Note that the impulse response function of water waves is never causal because of the dispersion of water waves\cite{falnes1995}.
However, such non-causality decreases as the distance $x_{\rm AP}$\cite{iida2022}, and the causality can be assumed in practice.
The form of the impulse response function of water waves and resultant prediction accuracy are described in the following subsections.

\subsection{Analytical form of impulse response function}
\begin{figure}[h]
\vspace*{0mm}
  \centerline{
\includegraphics[width=1.0\linewidth,clip]{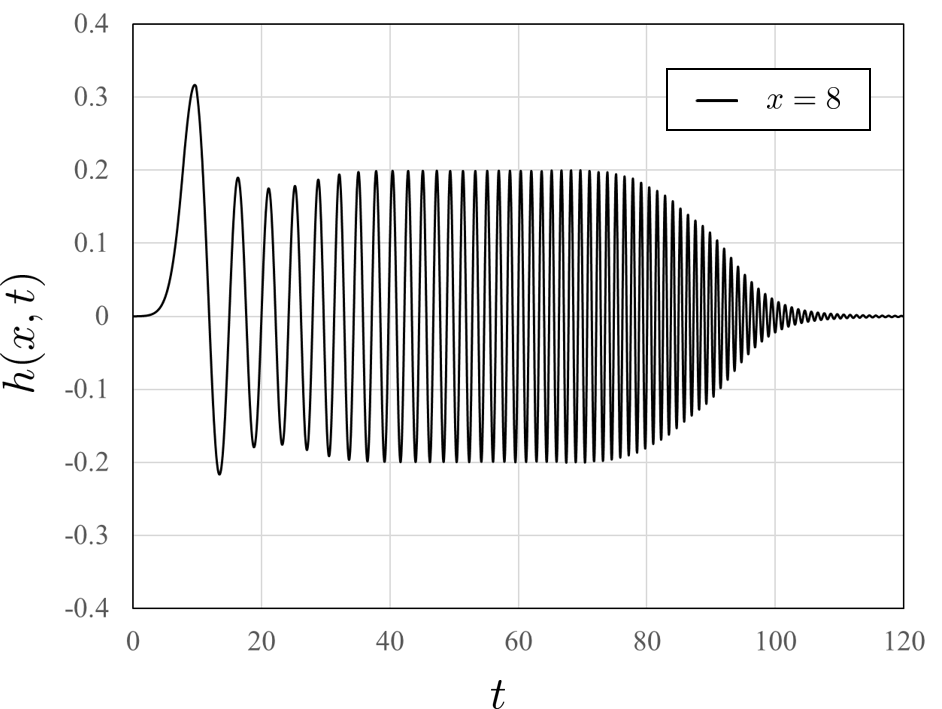}}
\par\vspace*{0mm}
\caption{Impulse response function of finite-depth water waves using the window function. In case of distance $x=8$, cutting-off frequencies $\Theta_1=\sqrt{20}$, and $\Theta_2=\sqrt{40}$ are plotted. Normalized values are shown.}
\label{fig2}
\end{figure}

The impulse response function is obtained by an inverse Fourier transform of a frequency response function $H(\omega)=\exp{(-ikx)}$ as
\begin{eqnarray}
\displaystyle h(x,t)=\frac{1}{2\pi}\int_{-\infty}^{\infty}e^{-ikx}e^{i\omega t}d\omega
\label{h:02}
\end{eqnarray}
where $\omega$ is a circular frequency and $k$ is a wave number.
Here, the wave number $k$ is bound by the dispersion relation.
The analytical solution of the impulse response function subject to deep water dispersion $k=\omega|\omega|$ throughout frequency is well-known\cite{korde2015, naito1986,davis1966,belmont2006}. 
However, this impulse response function overestimates a non-causality because of the dispersion relation\cite{iida2022}.
To obtain more accurate prediction results, the impulse response function using the dispersion relation of finite-depth water $\omega|\omega|=k\tanh{k}$ is examined\cite{iida2022,korde2020}.
As it is difficult to obtain a single solution valid for the whole time domain, the time domain is divided into three parts, and analytical solutions on respective time domains are obtained\cite{iida2022}.
In the paper, we also use the analytical solution of the impulse response function of the finite-depth water waves.
Note that the use of the dispersion of the finite-depth water waves is necessary even for predicting deep water waves because the impulse response function is obtained by an infinite range of frequencies.

The prediction using the impulse response function has a problem of the truncation error of the convolution integral as the impulse response function has an infinite length for the positive time.
The reference\cite{iida2022} used the finite impulse response (FIR) function for the analysis, but the low-pass filtered functions were also proposed based on the rectangular window function.
However, the rectangular window yields unnatural disturbance which may deteriorate the prediction accuracy.
Therefore, we derive a new form of the impulse response function using a new window function (cosine-type) when the time is large enough.
The new window function is defined as
\begin{eqnarray}
\displaystyle f(\omega)=
\left\{\begin{array}{lc}
1     & |\omega|<\Theta_1 \\
\displaystyle \cos{\bigg\{\frac{\pi}{2(\Theta_2-\Theta_1)}(|\omega|-\Theta_1)\bigg\}}  &\Theta_1\le |\omega|\le \Theta_2 \\
0     &\Theta_2<|\omega|
\end{array}\right.
\label{h:03}
\end{eqnarray}
where $\Theta_1$ and $\Theta_2$ are cutting-off frequencies to control the function.
Since the stationary wave number of the response function on the large time domain can assume the deep water dispersion, the impulse responses function is given as
\begin{eqnarray}
\displaystyle h(x,t)&=&\frac{1}{2\pi}\int_{-\infty}^{\infty}f(\omega)e^{-ikx}e^{i\omega t}d\omega\nonumber\\
\displaystyle &=&\frac{1}{\pi}{\rm Re}\bigg[\int_{0}^{\Theta_1}e^{i(\omega t-\omega^2 x)}d\omega\nonumber\\
\displaystyle &+&\int_{\Theta_1}^{\Theta_2}\cos{\bigg\{\frac{\pi}{2(\Theta_2-\Theta_1)}(\omega-\Theta_1)\bigg\}} e^{i(\omega t-\omega^2 x)}d\omega\bigg]\qquad
\label{h:04}
\end{eqnarray}
The first term on the right-hand side is the solution using the rectangular window function as follows\cite{iida2022}:
\begin{eqnarray}
\displaystyle &&\frac{1}{\pi}{\rm Re}\bigg[\int_{0}^{\Theta_1}e^{i(\omega t-\omega^2 x)}d\omega\bigg]\nonumber\\
&&=\frac{1}{\pi\sqrt{x}}\bigg[
\bigg\{
\mathcal{C}(\alpha_1)
+
\mathcal{C}(\beta_1)
\bigg\}\cos{\gamma_1}\nonumber\\
&&\quad \quad\quad+
\bigg\{
\mathcal{S}(\alpha_1)
+
\mathcal{S}(\beta_1)
\bigg\}\sin{\gamma_1}\bigg]
\label{h:05}
\end{eqnarray}
where $\mathcal{C}(\cdot)$ and $\mathcal{S}(\cdot)$ are the cosine and sine forms of the Fresnel integral\cite{abramowitz1964}, $\alpha_1=\sqrt{x}\{\Theta_1-t/(2x)\}$, $\beta_1=t/(2\sqrt{x})$, and $\gamma_1=t^2/(4x)$.
On the other hand, the second term is deformed as
\begin{eqnarray}
\displaystyle&& \frac{1}{\pi}{\rm Re}\bigg[\int_{\Theta_1}^{\Theta_2}\cos{\bigg\{\frac{\pi}{2(\Theta_2-\Theta_1)}(\omega-\Theta_1)\bigg\}} e^{i(\omega t-\omega^2 x)}d\omega\bigg]
\nonumber\\
&&=\sum_{i=2}^3\frac{1}{2\pi\sqrt{x}}\bigg[
\bigg\{
\mathcal{C}(\alpha_i)
+
\mathcal{C}(\beta_i)
\bigg\}\cos{\gamma_i}\nonumber\\
&&\quad\quad\quad\quad\quad+
\bigg\{
\mathcal{S}(\alpha_i)
+
\mathcal{S}(\beta_i)
\bigg\}\sin{\gamma_i}
\bigg]
\label{h:06}
\end{eqnarray}
where
\begin{eqnarray}
\displaystyle &&
\alpha_i=\left\{
\begin{array}{ll}
\displaystyle \sqrt{x}\bigg(\Theta_2-\frac{t}{2x}+\frac{\pi}{4x(\Theta_2-\Theta_1)}\bigg) &i=2\\
\displaystyle \sqrt{x}\bigg(\Theta_2-\frac{t}{2x}-\frac{\pi}{4x(\Theta_2-\Theta_1)}\bigg)&i=3
\end{array}\right.
\label{h:07}\\
\displaystyle &&
\beta_i=\left\{
\begin{array}{ll}
\displaystyle -\sqrt{x}\bigg(\Theta_1-\frac{t}{2x}+\frac{\pi}{4x(\Theta_2-\Theta_1)}\bigg) &i=2\\
\displaystyle -\sqrt{x}\bigg(\Theta_1-\frac{t}{2x}-\frac{\pi}{4x(\Theta_2-\Theta_1)}\bigg)&i=3
\end{array}\right.
\label{h:08}\\
\displaystyle &&
\gamma_i=\left\{
\begin{array}{ll}
\displaystyle 
\frac{1}{4x}\bigg(\frac{\pi}{2(\Theta_2-\Theta_1)}-t\bigg)^2+\frac{\pi\Theta_1}{2(\Theta_2-\Theta_1)}
&i=2\\
\displaystyle 
\frac{1}{4x}\bigg(\frac{\pi}{2(\Theta_2-\Theta_1)}+t\bigg)^2-\frac{\pi\Theta_1}{2(\Theta_2-\Theta_1)}
&i=3
\end{array}\right.\quad
\label{h:09}
\end{eqnarray}
The impulse response function with the filter (\ref{h:04}) is obtained by (\ref{h:05}) and (\ref{h:06}).

Combining the filtered impulse response function with analytical solutions\cite{iida2022}, the impulse response function on the whole time domain is given as
\begin{eqnarray}
\arraycolsep=2pt
\displaystyle 
&&h(x,t)=\nonumber\\
&&\left\{
\begin{array}{ll}
\vspace{1mm}
\displaystyle
\bigg(\frac{2}{x}\bigg)^{\frac{1}{3}}{\rm Ai}[\alpha_0]& t\le t_0\\
\vspace{1mm}
\displaystyle
\bigg(\frac{2}{t}\bigg)^{\frac{1}{3}}{\rm Ai}[\beta_0]
&t_0<t\le t_1\\
\vspace{1mm}
\displaystyle{\rm Re}\bigg[\frac{1}{\pi}c_g(k_0)\sqrt{\frac{2\pi}{t|\omega''(k_0)|}}e^{i[\omega(k_0)t-k_0x-\frac{\pi}{4}]}\bigg]&t_1< t\le t_2\\
\vspace{1mm}
\displaystyle
\sum_{i=1}^3\frac{1}{\pi\nu_i\sqrt{x}}\bigg[
\bigg\{
\mathcal{C}(\alpha_i)
+
\mathcal{C}(\beta_i)
\bigg\}\cos{\gamma_i}&\\
\quad\quad\quad\quad+
\bigg\{
\mathcal{S}(\alpha_i)
+
\mathcal{S}(\beta_i)
\bigg\}\sin{\gamma_i}
\bigg]&t_2<t
\end{array}\right. 
\label{h:10}
\end{eqnarray}
where ${\rm Ai}[\cdot]$ is the Airy function of the first kind, $\alpha_0=(x-t)(2/x)^{1/3}$, $\beta_0=(x-t)(2/t)^{1/3}$, $\nu_1=1$, $\nu_2=\nu_3=2$, $c_g$ is group velocity, $k_0$ is stationary wave number, and $\omega''$ is second-order derivative of frequency.
Besides, $t_0=x$, $t_1$ is the intersection point of the impulse response function on middle and large time domains, and $t_2$ is the time whose stationary phase is regarded as deep water waves.

The impulse response function calculated by (\ref{h:10}) is plotted in Fig. \ref{fig2}.
For cutting-off frequencies, $\Theta_1=\sqrt{20}$ and $\Theta_2=\sqrt{40}$ are used.
Figure \ref{fig2} shows the case of $x=8$, where the distance is sufficiently long to assume causality.
When the time is large enough, the response function attenuates to zero because of the filtering by the window function.
As a result, the concern of the truncation error is avoidable.

\subsection{Prediction error}

\begin{figure}[h]
\vspace*{0mm}
  \centerline{
\includegraphics[width=1.0\linewidth,clip]{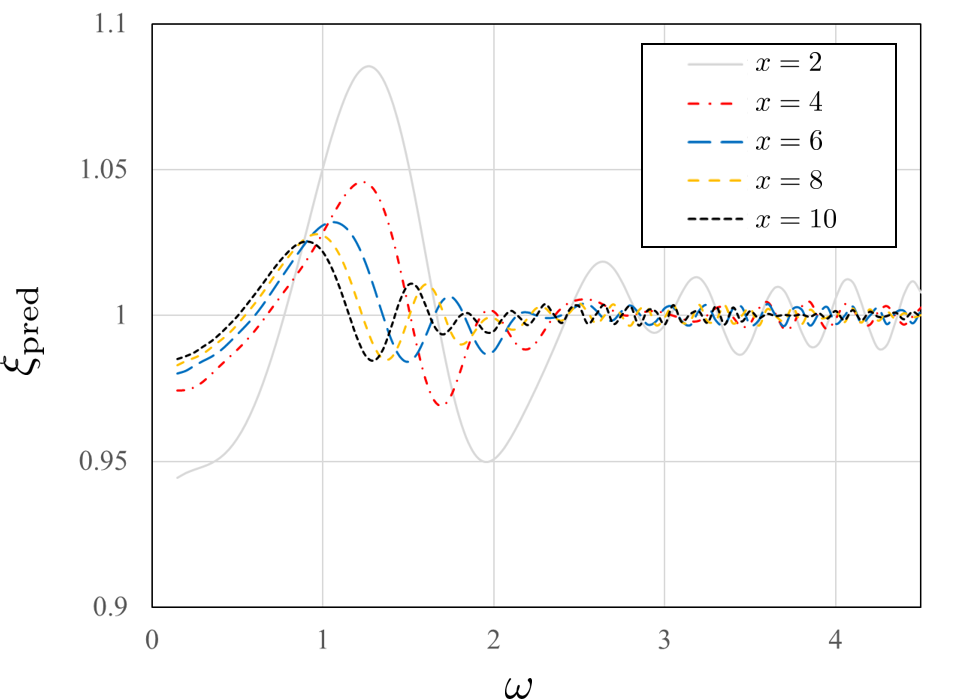}}
\par\vspace*{0mm}
\caption{Relation between predicted wave amplitude $\xi_{\rm pred}$ and frequency $\omega$. Results are plotted using different distances $x$. Values are normalized.}
\label{fig3}
\end{figure}

It is important to investigate the prediction accuracy, and this subsection discusses prediction errors.
Prediction based on the impulse response function contains some errors due to mathematical formulation (linear potential theory), non-causality, truncation up to finite length, discretization, and so on.
The error of the assumption of causal function is estimated by the factor of non-causality\cite{iida2022}.
In the case of the finite-depth water waves, the factor of non-causality is given as
\begin{eqnarray}
\displaystyle \mathcal{F}(x)=\frac{h(x,0)}{h(x,t_0)}=\frac{{\rm Ai}\left[(2x^2)^{\frac{1}{3}}\right]}{{\rm Ai}\left[0\right]}
\label{e:01}
\end{eqnarray}
Equation (\ref{e:01}) indicates that the non-causal effect exponentially decays as the distance, and the factor is less than 0.01 (1\%) when the distance is $x\ge4.4$.
Note that this factor is not an error of the prediction but the error of ignoring the non-causal range.

In the paper, the prediction accuracy of sinusoidal waves with respect to frequency is investigated.
Using the impulse response function method, the phase of sinusoidal waves can be precisely predicted.
On the other hand, the prediction accuracy of the amplitude depends on the target frequency.
Therefore, the predicted wave amplitude $\xi_{\rm pred}$ is calculated to frequency $\omega$ as shown in Fig. \ref{fig3}.
Predicted wave amplitude is normalized by the amplitude of input waves.
For the calculation of the impulse response function, $\Theta_1=\sqrt{20}$ and $\Theta_2=\sqrt{40}$ are used.
Figure \ref{fig3} includes results of $x=2, 4, 6, 8$ and 10.
The exact solution is $\xi_{\rm pred}=1$, but the results show a small discrepancy from the exact value.
This discrepancy decreases as the distance.
The normalized frequency is identified by the dispersion: shallow water ($\omega<\pi/10$), finite-depth water ($\pi/10 \le \omega <\sqrt{\pi}$), and deep water ($\sqrt{\pi}\le\omega$).
When the target frequency is in the deep water regime, the error of the predicted amplitude is less than 0.01 (1\%) except for $x=2$.
When the frequency is in the finite-depth water regime, on the other hand, the error becomes 0.01 to 0.05 (1 to 5\%).
When the frequency is in the shallow water regime, the error converges to values as the frequency is small.
The converged value is smaller than the correct one, and this gap becomes small as the distance is long.
Summarizing above, the prediction of finite-depth water waves offers a longer distance than deep water waves.

\section{Extension to bi-directional water waves}

\subsection{Decomposition and prediction methods}
The decomposition method of bi-directional waves is formulated using measured surface elevation data at points A and B.
Here, the integral range of (\ref{h:01}) is truncated up to the finite length $[0, T_{\rm max}]$, and the integral is discretized by the rectangular approximation as
\begin{eqnarray}
\displaystyle \xi_{\rm P}(x_{\rm AP},t)&\approx& \sum_{k=0}^Kh(x_{\rm AP},k\Delta t)\xi_{\rm A}(t-k\Delta t)\Delta t\nonumber\\
\displaystyle &\equiv&\sum_{k=0}^Kb^{\rm (AP)}_k \xi_{\rm A}(t-k\Delta t)
\label{bi:01}
\end{eqnarray}
where $b^{\rm (AP)}_k=h(x_{\rm AP},k\Delta t)\Delta t$, and $\Delta t=T_{\rm max}/K$ is a time step size.

\begin{table}[h]
 \caption{Definitions of waves at points A, B, and P. Progressive waves at point A ($\eta_{\rm A}$) are used to predict progressive waves at points P and B ($\zeta_{\rm AP}$ and $\zeta_{\rm AB}$). Similarly, Regressive waves at point B ($\eta_{\rm B}$) are used to predict regressive waves at points P and A ($\zeta_{\rm BP}$ and $\zeta_{\rm BA}$). Actual bi-directional waves are represented by the superposition of progressive and regressive waves, i.e. $\xi_{\rm A}=\eta_{\rm A}+\zeta_{\rm BA}$, $\xi_{\rm B}=\zeta_{\rm AB}+\eta_{\rm B}$, and $\xi_{\rm P}=\zeta_{\rm AP}+\zeta_{\rm BP}$.}
 \label{table1}
 \par\vspace*{2mm}
 \centering
\begin{tabular}{ |c|c|c|c| } 
\hline
&A&P&B \\
\hline
Progressive &$\eta_{\rm A}(t)$ &$\zeta_{\rm AP}(t)$&$\zeta_{\rm AB}(t)$ \\ 
Regressive  &$\zeta_{\rm BA}(t)$ & $\zeta_{\rm BP}(t)$&$\eta_{\rm B}(t)$ \\ 
Bi-directional&$\xi_{\rm A}(t)$&$\xi_{\rm P}(t)$&$\xi_{\rm B}(t)$\\
\hline
\end{tabular}
\end{table}
To formulate the decomposition method, wave components at each point are defined in Table \ref{table1}.
We consider progressive and regressive waves, respectively.
Progressive waves at up-stream point (i.e. A) are defined as $\eta_{\rm A}$.
Progressive waves at down-stream points (i.e. P and B), defined as $\zeta_{\rm AP}$ and $\zeta_{\rm AB}$, are predicted using $\eta_{\rm A}$.
Similar to them, regressive waves at up-stream point B are identified by $\eta_{\rm B}$, and predicted regressive waves at A and P are $\zeta_{\rm BA}$ and $\zeta_{\rm BP}$.
Bi-directional waves are represented by the superposition of progressive and regressive waves as $\xi_{\rm A}=\eta_{\rm A}+\zeta_{\rm BA}$, $\xi_{\rm B}=\zeta_{\rm AB}+\eta_{\rm B}$, and $\xi_{\rm P}=\zeta_{\rm AP}+\zeta_{\rm BP}$.
Here, $\xi_{\rm A}$, $\xi_{\rm P}$, and $\xi_{\rm B}$ are actually measured wave data using wave-rider buoys/wave gauges.
These are expressed as
\begin{eqnarray}
\displaystyle \xi_{\rm A}(t)=\eta_{\rm A}(t)+\zeta_{\rm BA}(t)=\eta_{\rm A}(t)+\sum_{k=0}^Kb^{\rm (BA)}_k \eta_{\rm B}(t-k\Delta t)
\label{bi:02}\\
\displaystyle \xi_{\rm B}(t)=\eta_{\rm B}(t)+\zeta_{\rm AB}(t)=\eta_{\rm B}(t)+\sum_{k=0}^Kb^{\rm (AB)}_k \eta_{\rm A}(t-k\Delta t)
\label{bi:03}
\end{eqnarray}
Here $b_k^{(\rm AB)}=b_k^{(\rm BA)}$.
This is an initial-value problem, and unknown parameters $\eta_{\rm A}(t)$ and $\eta_{\rm B}(t)$ are given from (\ref{bi:02}) and (\ref{bi:03}) as
\begin{eqnarray}
\displaystyle \left[\begin{array}{l}
\eta_A(t)\\
\eta_B(t)
\end{array}\right]
=
 \frac{1}{1-b_0^{({\rm AB}),2}}
\left[\begin{array}{cc}
1&-b_0^{(\rm AB)}\\
-b_0^{(\rm AB)}&1
\end{array}\right]
\left[\begin{array}{l}
\xi_A(t)\\
\xi_B(t)
\end{array}\right]\nonumber\\
\displaystyle - \frac{1}{1-b_0^{({\rm AB}),2}}
\left[\begin{array}{cc}
1&-b_0^{(\rm AB)}\\
-b_0^{(\rm AB)}&1
\end{array}\right]
\left[\begin{array}{l}
\sum_{k=1}^Kb_k^{(\rm AB)}\eta_B(t-k\Delta t)\\
\sum_{k=1}^Kb_k^{(\rm AB)}\eta_A(t-k\Delta t)
\end{array}\right]
\label{bi:04}
\end{eqnarray}
We further assume causality at the present time as $b_0^{({\rm AB}),2}\sim 0$.
It is worth noting that this assumption is not required for the calculation here, but this assumption enables the extension to a multi-directional prediction (i.e. three-dimensional problem).
Then, (\ref{bi:04}) is simplified as
\begin{eqnarray}
\displaystyle \left[\begin{array}{l}
\eta_A(t)\\
\eta_B(t)
\end{array}\right]
=
\left[\begin{array}{cc}
1&-b_0^{(\rm AB)}\\
-b_0^{(\rm AB)}&1
\end{array}\right]
\left[\begin{array}{l}
\xi_A(t)\\
\xi_B(t)
\end{array}\right]\nonumber\\
\displaystyle - 
\left[\begin{array}{cc}
1&-b_0^{(\rm AB)}\\
-b_0^{(\rm AB)}&1
\end{array}\right]
\left[\begin{array}{l}
\sum_{k=1}^Kb_k^{(\rm AB)}\eta_B(t-k\Delta t)\\
\sum_{k=1}^Kb_k^{(\rm AB)}\eta_A(t-k\Delta t)
\end{array}\right]
\label{bi:05}
\end{eqnarray}
Equation (\ref{bi:05}) indicates that the decomposed components at the present time ($\eta_{\rm A}(t)$ and $\eta_{\rm B}(t)$) can be calculated by measured wave data ($\xi_{\rm A}(t)$ and $\xi_{\rm B}(t)$) and the decomposed components at past times ($\eta_{\rm A}(t-k\Delta t)$ and $\eta_{\rm B}(t-k\Delta t)$, $k=1,2,\cdots,K$) which are already known.
As this is a time-developing problem, the initial values are necessary.

Once bi-directional waves are decomposed, the free surface elevation at point P (between points A and B) is predicted by
\begin{eqnarray}
\displaystyle \xi_{\rm P}(t)=\sum_{k=0}^Kb^{\rm (AP)}_k \eta_{\rm A}(t-k\Delta t)+\sum_{k=0}^Kb^{\rm (BP)}_k \eta_{\rm B}(t-k\Delta t)
\label{bi:06}
\end{eqnarray}

\begin{figure*}[h]
\vspace*{0mm}
  \centerline{
\includegraphics[width=1.0\linewidth,clip]{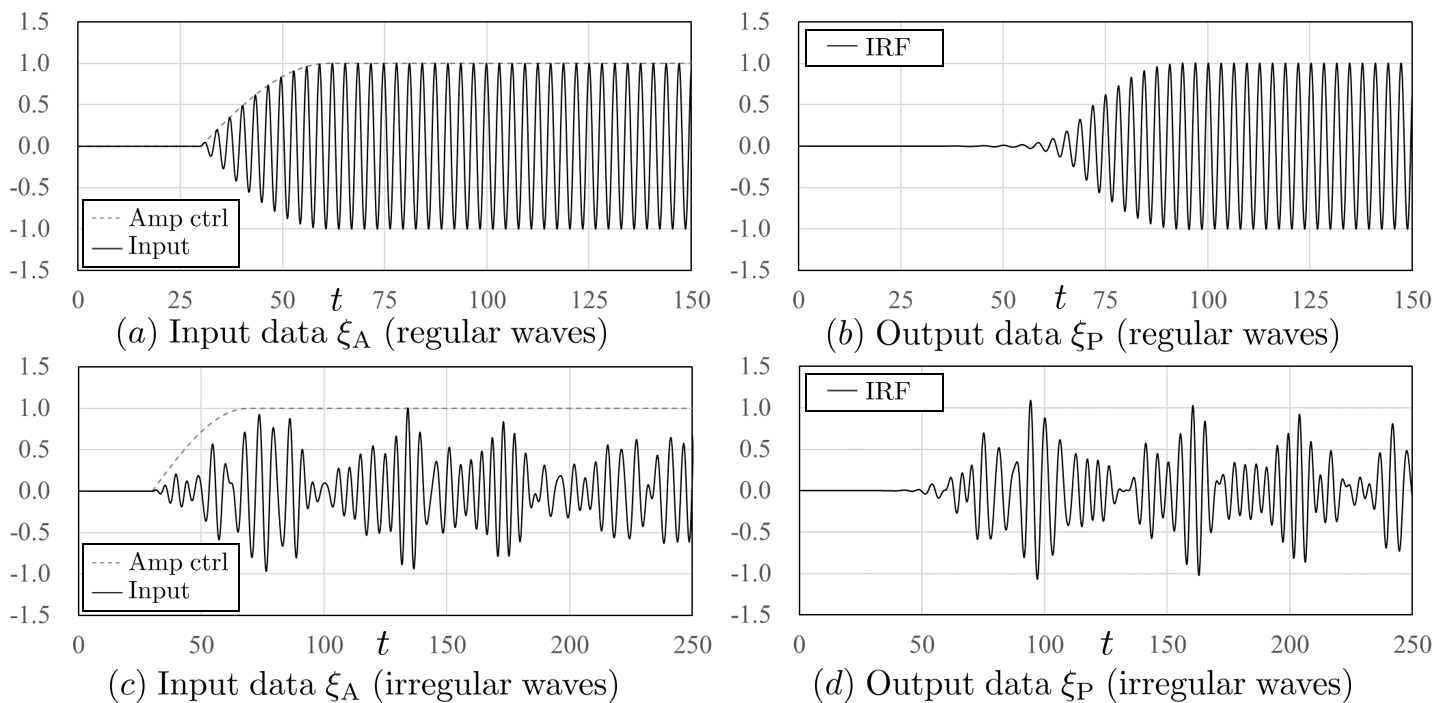}}
\par\vspace*{0mm}
\caption{Generations of initial uniform wave train. Figures above are case of regular waves ($\omega=2.0$), and figures below are case of irregular waves (Pierson-Moskowitz, $T_{1}=2\pi/1.6$ and $H_{1/3}=2.0$).
$(a)$ and $(c)$ are input wave profiles at point A (inputs of virtual wave maker). 
Startup amplitudes are controlled to gradually approach a predetermined value. 
$(b)$ and $(d)$ are free surface elevation at down-stream point P where $x_{\rm AP}=8$. Normalized values are shown.}
\label{fig4}
\end{figure*}
\subsection{State-space representation}
Since many control methods of the WEC are based on state-space models (e.g. \cite{xuhui2019,gaebele2020}), it may be convenient if the proposed method is also represented by the state-space model.
In addition, the computational speed of the state-space model is faster than that of the convolution integral\cite{taghipour2008,zhang2015}.
Therefore, this subsection describes the state-space representation of (\ref{bi:05}) and (\ref{bi:06}).

Firstly, the state-space representation of (\ref{bi:05}) is given as
\begin{eqnarray}
\displaystyle 
&&\mbox{\boldmath $x$}_{\rm AB}(t)=\mbox{\boldmath $A$}\mbox{\boldmath $x$}_{\rm AB}(t-\Delta t)+\mbox{\boldmath $B$}\mbox{\boldmath $\xi$}_{\rm AB}(t-\Delta t)
\label{ss:01}\\
\displaystyle 
&&\mbox{\boldmath $\eta$}_{\rm AB}(t)=\mbox{\boldmath $C$}\mbox{\boldmath $x$}_{\rm AB}(t)+\mbox{\boldmath $D$}\mbox{\boldmath $\xi$}_{\rm AB}(t)
\label{ss:02}
\end{eqnarray}
where
\begin{eqnarray}
\displaystyle 
\mbox{\boldmath $\eta$}_{\rm AB}(t)=\left[\begin{array}{l}
\eta_{\rm A}(t)\\
\eta_{\rm B}(t)
\end{array}\right],\;
\mbox{\boldmath $\xi$}_{\rm AB}(t)=\left[\begin{array}{l}
\xi_{\rm A}(t)\\
\xi_{\rm B}(t)
\end{array}\right]
\label{ss:03}
\end{eqnarray}
\begin{eqnarray}
\arraycolsep=2pt
\displaystyle 
\mbox{\boldmath $A$}=
\left[\begin{array}{cccccc|ccccccccc}
0&0&\cdots&0&0&b_0^{({\rm AB})}b_K^{({\rm AB})}&0&\cdots&\cdots&\cdots&0&-b_K^{({\rm AB})}\\
1&0&\cdots&0&0&b_0^{({\rm AB})}b_{K-1}^{({\rm AB})}&\vdots&&&&\vdots&-b_{K-1}^{({\rm AB})}\\
\vdots&\ddots&&&\vdots&\vdots&\vdots&&&&\vdots&\vdots\\
\vdots&&\ddots&&\vdots&\vdots&\vdots&&&&\vdots&\vdots\\
0&0&&1&0&b_0^{({\rm AB})}b_2^{({\rm AB})}&\vdots&&&&\vdots&-b_2^{({\rm AB})}\\
0&0&\cdots&0&1&b_0^{({\rm AB})}b_1^{({\rm AB})}&0&\cdots&\cdots&\cdots&0&-b_1^{({\rm AB})}\\
\hline
0&\cdots&\cdots&\cdots&0&-b_K^{({\rm AB})}&0&0&\cdots&0&0&b_0^{({\rm AB})}b_K^{({\rm AB})}\\
\vdots&&&&\vdots&-b_{K-1}^{({\rm AB})}&1&0&\cdots&0&0&b_0^{({\rm AB})}b_{K-1}^{({\rm AB})}\\
\vdots&&&&\vdots&\vdots&\vdots&\ddots&&&\vdots&\vdots\\
\vdots&&&&\vdots&\vdots&\vdots&&\ddots&&\vdots&\vdots\\
\vdots&&&&\vdots&-b_2^{({\rm AB})}&0&0&&1&0&b_0^{({\rm AB})}b_2^{({\rm AB})}\\
0&\cdots&\cdots&\cdots&0&-b_1^{({\rm AB})}&0&0&\cdots&0&1&b_0^{({\rm AB})}b_1^{({\rm AB})}\\
\end{array}\right]
\label{ss:04}
\end{eqnarray}
\begin{eqnarray}
\arraycolsep=2pt
\displaystyle 
\mbox{\boldmath $B$}=
\left[\begin{array}{cc}
2b_0^{({\rm AB})}b_K^{({\rm AB})}&-b_K^{({\rm AB})}\\
2b_0^{({\rm AB})}b_{K-1}^{({\rm AB})}&-b_{K-1}^{({\rm AB})}\\
\vdots&\vdots\\
\vdots&\vdots\\
2b_0^{({\rm AB})}b_2^{({\rm AB})}&-b_2^{({\rm AB})}\\
2b_0^{({\rm AB})}b_1^{({\rm AB})}&-b_1^{({\rm AB})}\\
\hline
-b_K^{({\rm AB})}&2b_0^{({\rm AB})}b_K^{({\rm AB})}\\
-b_{K-1}^{({\rm AB})}&2b_0^{({\rm AB})}b_{K-1}^{({\rm AB})}\\
\vdots&\vdots\\
\vdots&\vdots\\
-b_2^{({\rm AB})}&2b_0^{({\rm AB})}b_2^{({\rm AB})}\\
-b_1^{({\rm AB})}&2b_0^{({\rm AB})}b_1^{({\rm AB})}\\
\end{array}\right],\;
\mbox{\boldmath $D$}=\left[\begin{array}{cc}
1&-b_0^{({\rm AB})}\\
-b_0^{({\rm AB})}&1
\end{array}\right]
\label{ss:05}
\end{eqnarray}
\begin{eqnarray}
\arraycolsep=2pt
\displaystyle 
\mbox{\boldmath $C$}=\left[\begin{array}{cccccc|cccccc}
0&\cdots&\cdots&\cdots&0&1&0&\cdots&\cdots&\cdots&\cdots&0\\
0&\cdots&\cdots&\cdots&\cdots&0&0&\cdots&\cdots&\cdots&0&1
\end{array}\right]
\label{ss:06}
\end{eqnarray}
Here, $\mbox{\boldmath $x$}_{\rm AB}(t)$ is the state vector, $\mbox{\boldmath $A$}$ is the state matrix, $\mbox{\boldmath $B$}$ is the input matrix, $\mbox{\boldmath $C$}$ is the output matrix, and $\mbox{\boldmath $D$}$ is the feedthrough matrix of this system.
The state vector $\mbox{\boldmath $x$}_{\rm AB}(t)$ saves decomposed waves in the past.
On the other hand, the system to predict waves at point P, i.e. (\ref{bi:06}), is deformed as
\begin{eqnarray}
\displaystyle 
&&\mbox{\boldmath $x$}_{\rm P}(t)=\mbox{\boldmath $E$}\mbox{\boldmath $x$}_{\rm P}(t-\Delta t)+\mbox{\boldmath $F$}\mbox{\boldmath $\eta$}_{\rm AB}(t-\Delta t)
\label{ss:07}\\
\displaystyle 
&&\xi_{\rm P}(t)=\mbox{\boldmath $G$}\mbox{\boldmath $x$}_{\rm P}(t)+\mbox{\boldmath $H$}\mbox{\boldmath $\eta$}_{\rm AB}(t)
\label{ss:08}
\end{eqnarray}
where
\begin{eqnarray}
\arraycolsep=2pt
\displaystyle 
\mbox{\boldmath $E$}=
\left[\begin{array}{cccccc}
0&0&\cdots&\cdots&0&0\\
1&0&&&0&0\\
\vdots&\ddots&&&\vdots&\vdots\\
\vdots&&\ddots&&\vdots&\vdots\\
0&0&&\ddots&0&0\\
0&0&\cdots&\cdots&1&0\\
\end{array}\right],\;
\mbox{\boldmath $F$}=
\left[\begin{array}{cc}
b^{({\rm AP})}_K&b^{({\rm BP})}_K\\
b^{({\rm AP})}_{K-1}&b^{({\rm BP})}_{K-1}\\
\vdots&\vdots\\
\vdots&\vdots\\
b^{({\rm AP})}_2&b^{({\rm BP})}_2\\
b^{({\rm AP})}_1&b^{({\rm BP})}_1\\
\end{array}\right]
\label{ss:09}
\end{eqnarray}
\begin{eqnarray}
\arraycolsep=2pt
\displaystyle 
\mbox{\boldmath $G$}=\left[\begin{array}{cccccc}
0&\cdots&\cdots&\cdots&0&1
\end{array}\right]
\label{ss:10}
\end{eqnarray}
\begin{eqnarray}
\arraycolsep=2pt
\displaystyle 
\mbox{\boldmath $H$}=\left[\begin{array}{cc}
b^{({\rm AP})}_0&b^{({\rm BP})}_0
\end{array}\right]
\label{ss:11}
\end{eqnarray}
Note that $\mbox{\boldmath $x$}_{\rm P}(t)$ is the state vector, $\mbox{\boldmath $E$}$ is the state matrix, $\mbox{\boldmath $F$}$ is the input matrix, $\mbox{\boldmath $G$}$ is the output vector, and $\mbox{\boldmath $H$}$ is the feedthrough vector of this system.

Combining two systems (\ref{ss:01}), (\ref{ss:02}), (\ref{ss:07}), and (\ref{ss:08}), the new system to predict waves at point P using the wave data at points A and B is obtained as
\begin{eqnarray}
\arraycolsep=2pt
\displaystyle &&
\mbox{\boldmath $x$}(t)=
\left[\begin{array}{c|c}
\mbox{\boldmath $A$}&\mbox{\boldmath $O$}\\
\hline
\mbox{\boldmath $FC$}&\mbox{\boldmath $E$}
\end{array}\right]
\mbox{\boldmath $x$}(t-\Delta t)+
\left[\begin{array}{c}
\mbox{\boldmath $B$}\\
\hline
\mbox{\boldmath $FD$}
\end{array}\right]\mbox{\boldmath $\xi$}_{\rm AB}(t-\Delta t)
\label{ss:12}\\
\displaystyle && \xi_{\rm P}(t)=
\left[\begin{array}{c|c}
\mbox{\boldmath $HC$}&\mbox{\boldmath $G$}
\end{array}\right]
\mbox{\boldmath $x$}(t)+\mbox{\boldmath $HD$}\mbox{\boldmath $\xi$}_{\rm AB}(t)
\label{ss:13}
\end{eqnarray}
where
\begin{eqnarray}
\arraycolsep=2pt
\displaystyle &&
\mbox{\boldmath $x$}(t)=
\left[\begin{array}{c}
\mbox{\boldmath $x$}_{\rm AB}(t)\\
\hline
\mbox{\boldmath $x$}_{\rm P}(t)
\end{array}\right]
\label{ss:14}
\end{eqnarray}

\begin{figure*}[t]
\vspace*{0mm}
  \centerline{
\includegraphics[width=1.0\linewidth,clip]{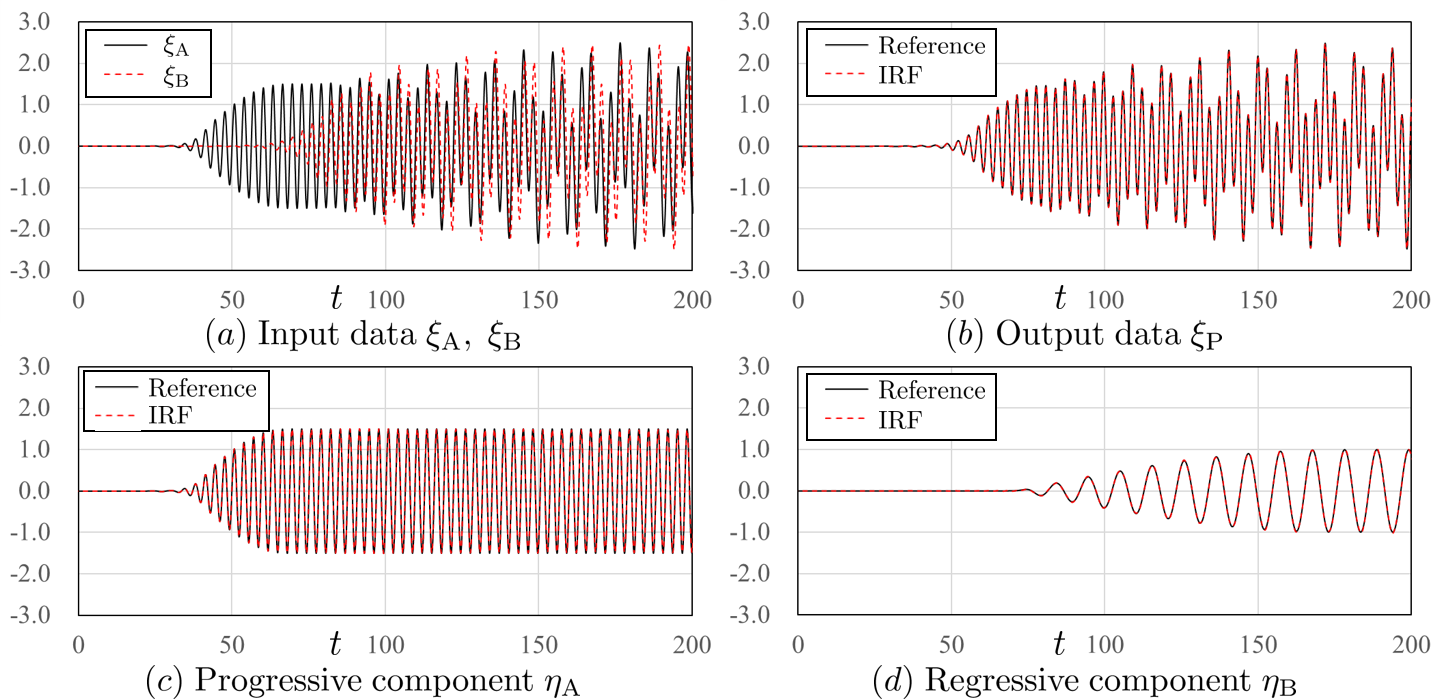}}
\par\vspace*{0mm}
\caption{Prediction of initial uniform bi-directional waves consisting of two regular waves. (a) Free surface elevation at point A and B. These wave profiles are inputs of prediction. (b) Free surface elevation at point P predicted by impulse response function (IRF). (c) Progressive component of bi-directional waves at point A of which frequency is $\omega=2.0$. (d) Regressive component of bi-directional waves at point B of which frequency is $\omega=0.6$. Normalized values are shown. }
\label{fig5}
\end{figure*}

\section{Generation of initial uniform wave train}

One of the advantages of the impulse response function method is its applicability to the prediction of initial uniform waves/transient waves.
As the prediction based on the Fourier coefficients estimation assumes an infinite length of sinusoidal waves, this method could be used for the prediction of such waves.
Especially, the prediction accuracy of the wavefront is not ensured due to the Gibbs phenomenon.
Therefore, in the paper, initial uniform waves are considered to emphasize the advantage of the impulse response function method.
In this section, a generation method of the initial uniform wave train is briefly explained.

To generate waves using an arbitrary wave profile, the uni-directional wave prediction based on the impulse response function is used.
We consider two points A and P, where A is located at the up-stream of P.
Point A is regarded as a virtual wave maker, and an arbitrary wave profile is set as input of the motion of the wave maker.
Then, the free surface elevation at point P is predicted by (\ref{e:01}).
Using this process, the wavefront can be reproduced.

Examples of initial uniform wave train generation are shown in Fig. \ref{fig4}.
The figures above are the case of regular waves whose frequency is $\omega=2.0$.
On the other hand, the figures below are the case of irregular waves.
Pierson-Moskowitz type spectrum is used where a mean wave period is $T_1=2\pi/1.6$ and the wave amplitude is normalized by significant wave amplitude $H_{1/3}/2$.
Figures 4 $(a)$ and $(c)$ are input wave profiles of the virtual wave maker located at point A.
The wave maker starts its generation at $t=30$.
The startup amplitude is controlled to gradually approach a predetermined value.
This amplitude control is shown in the figures.
Using these input profiles, wave elevation at an arbitrary down-stream point is calculated.
Free surface elevations at point P with $x_{\rm AP}=8$ are shown in Fig.\ref{fig4} $(b)$ and $(d)$.
As these wave profiles are initial uniform, it is difficult to reproduce the wavefront using the Fourier series expansion.
On the other hand, the method of the impulse response function can realize such a wave propagation process, and the initial uniform wave train can be generated.

\section{Results and discussion}
\subsection{Numerical validations}

The proposed decomposition and prediction methods are validated by numerical simulations.
For the first simulation, we consider bi-directional waves consisting of initial uniform regular waves. 
Using the generation method described in the previous section, two initial uniform uni-directional regular waves are prepared.
The first waves propagate from the negative $x$ direction (i.e. progressive waves), and the second waves propagate from the positive $x$ direction (i.e. regressive waves).
Then, the bi-directional wave field is calculated by the superposition of these two waves.

For the simulation setup, three points A, B, and P are arranged with the distances $x_{\rm AP}=4$ and $x_{\rm AB}=8$.
Cutting-off frequencies are set as $\Theta_1=\sqrt{20}$ and $\Theta_2=\sqrt{40}$.
Besides, $\Delta t=0.01\sqrt{9.81/0.45}$ and $T_{\rm max}=30\sqrt{9.81/0.45}$ are used where $\sqrt{9.81/0.45}(\approx4.67)=\sqrt{g/d}$ is square root of gravitational acceleration/water depth that normalizes time.
The frequency of the first waves (progressive component) is $\omega=2.0$, and that of the second waves (regressive component) is $\omega=0.6$.
Each wave component is generated by the virtual wave maker located in front of the up-stream point with distance $x=4$. 
Figure \ref{fig5} (a) shows free surface elevations at points A and B.
Using these inputs, the free surface elevation at point P is predicted as shown in Fig. \ref{fig5} (b).
The reference indicates the elevation calculated by the superposition of two uni-directional waves.
Decomposed progressive and regressive wave components are also shown in Fig. \ref{fig5} (c) and (d).
Results of the impulse response function show perfect agreement with the references, and progressive and regressive waves are well decomposed.
As a result, the wave elevation at point P is also predictable with high accuracy as in Fig. \ref{fig5} (b).

\begin{figure*}[t]
\vspace*{0mm}
  \centerline{
\includegraphics[width=1.0\linewidth,clip]{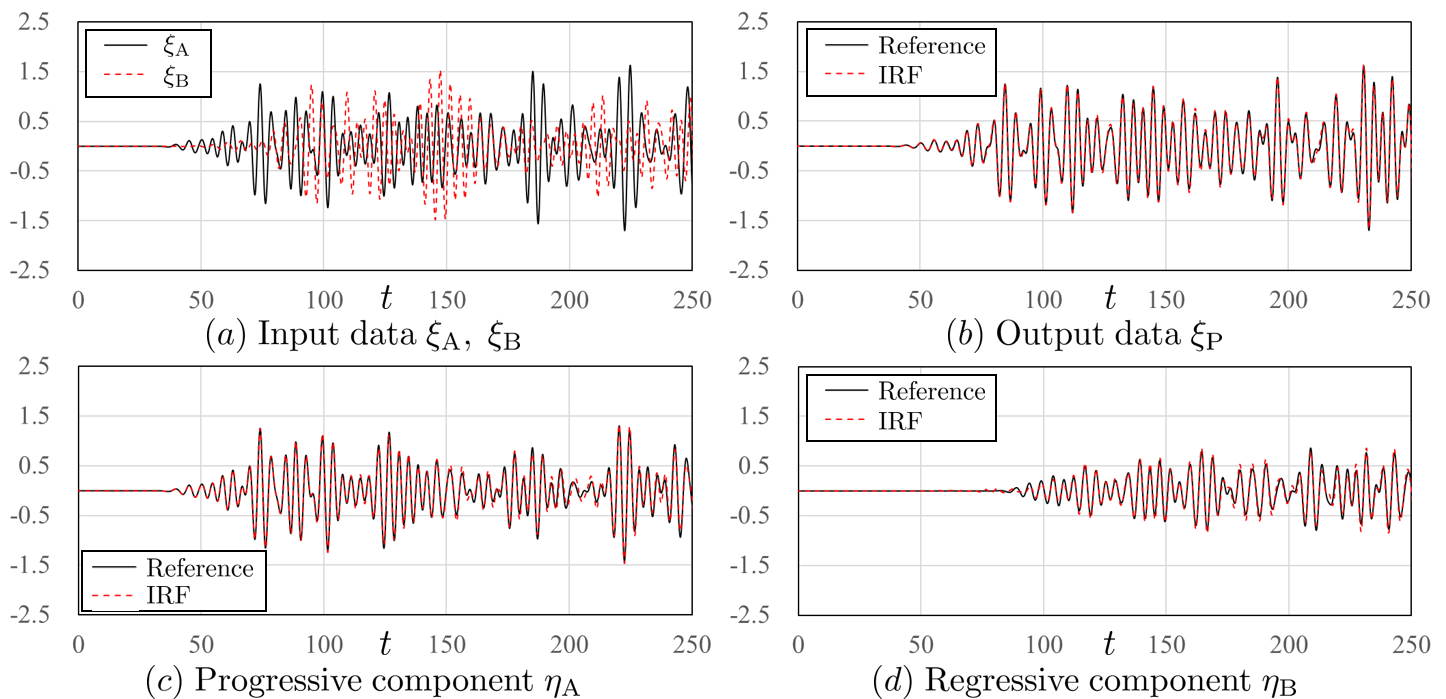}}
\par\vspace*{0mm}
\caption{Prediction of initial uniform bi-directional waves consisting of irregular waves. Irregular waves are based on Pierson-Moskowitz type spectrum, and mean wave period is $T_1=2\pi/1.8$. (a) Free surface elevation at points A and B. These wave profiles are inputs of prediction. (b) Free surface elevation at point P predicted by impulse response function (IRF). (c) Progressive component of bi-directional waves at point A. (d) Regressive component of bi-directional waves at point B. Normalized values are shown. }
\label{fig6}
\end{figure*}

Secondly, we demonstrate the prediction of bi-directional irregular waves.
The simulation setup is the same as the regular wave case.
Irregular waves are based on the Pierson-Moskowitz type spectrum.
A mean wave period of both progressive and regressive waves is $T_1=2\pi/1.8$, and wave amplitude is normalized by significant wave amplitude.
Similar to the regular wave case, Fig. \ref{fig6} (a) is the free surface elevations at points A and B, (b) is the surface elevation at point P predicted by the impulse response function method, and (c) and (d) are progressive and regressive components of bi-directional waves decomposed by the proposed method, respectively.
Predicted results show a slight difference in amplitude from references.
According to Fig. \ref{fig3}, the prediction error depends on frequency.
As irregular waves are consisting of a range of frequencies, the prediction contains a slight error.
However, the overall agreement is sufficiently good.

\subsection{Experimental validation}

Experimental validation is also carried out to demonstrate the applicability of the proposed method.
Experimental wave data were measured in a two-dimensional water tank at Osaka University, Japan, where the tank length is 14 m, and the width is 0.3 m.
The tank was filled with pure water, and the water depth was 0.45 m.
To measure free surface elevations, supersonic wave gauges were used, and the sampling time of gauges was 0.01 s.
Three wave gauges were arranged with distances $x_{\rm AP}=1.5/0.45=3.3$ and $x_{\rm AB}=3.0/0.45=6.7$. 
The plunger-type wave maker generates progressive waves at one end of the tank.
These waves are reflected at another end, becoming regressive waves.

\begin{figure*}[t]
\vspace*{0mm}
  \centerline{
\includegraphics[width=1.0\linewidth,clip]{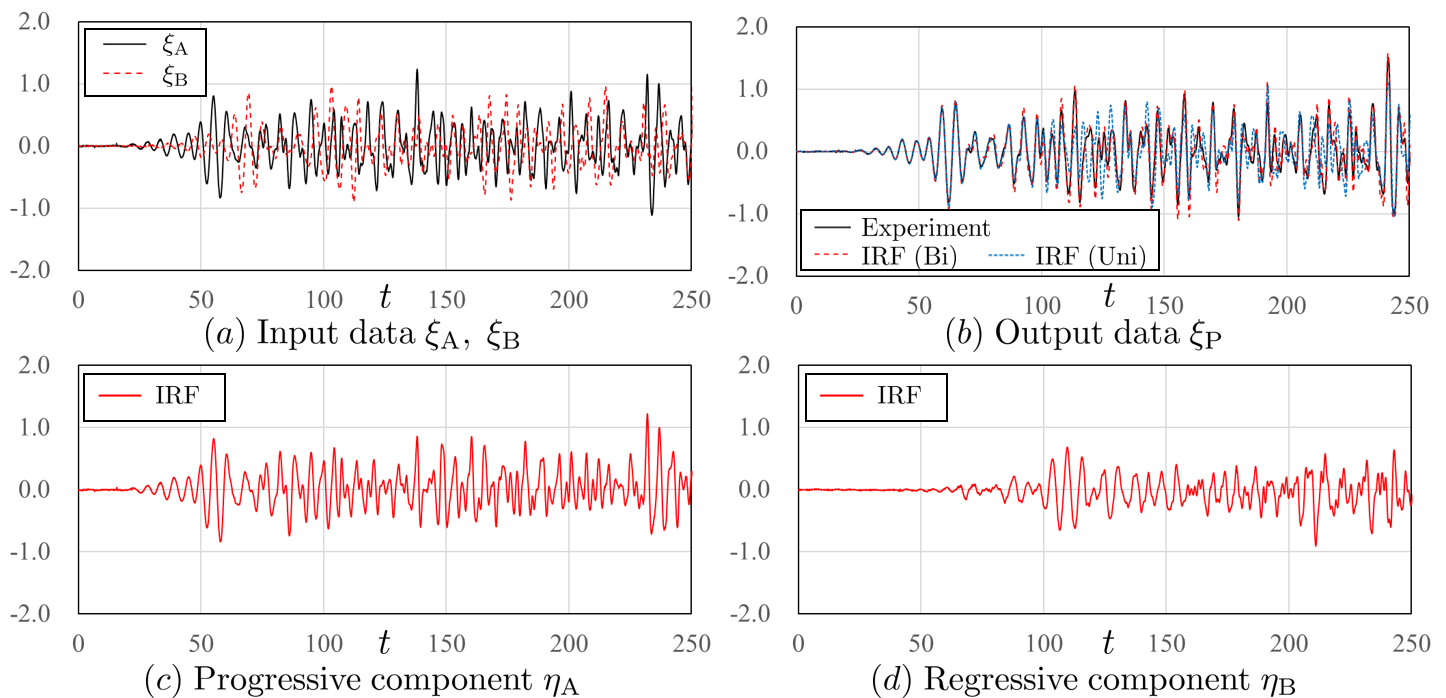}}
\par\vspace*{0mm}
\caption{Prediction of initial uniform bi-directional irregular waves using measured wave profiles in tank experiment. Irregular waves are based on JONSWAP-type spectrum, and significant wave period is $T{1/3}=5.6$. (a) Free surface elevation at points A and B obtained by experiment. These wave profiles are inputs of prediction. (b) Free surface elevation at point P predicted by impulse response function (IRF). 
Prediction using input of two points A and B is denoted as IRF (Bi), and prediction using input of only point A is IRF (Uni).
(c) Progressive component of bi-directional waves at point A. (d) Regressive component of bi-directional waves at point B. Normalized values are shown. }
\label{fig7}
\end{figure*}

Here, irregular waves based on the JONSWAP spectrum\cite{stansberg2002} with $\gamma=3.3$ were considered.
The significant wave period $T_{1/3}=1.2\sqrt{9.81/0.45}(\approx5.60)$ is used; the significant frequency is $\omega_{1/3}=1.1$.
For the prediction, the same numerical conditions are used with the numerical validations in subsection 6.1.
Figure \ref{fig7} (a) is measured wave profiles at points A and B, (b) is the free surface elevation at point P, and (c) and (d) are predicted progressive and regressive wave components.
Figure \ref{fig7} (b) contains the experimental result, numerical result using the bi-directional prediction (i.e. (\ref{bi:06}) and inputs of points A and B are used), and numerical result using the uni-directional prediction (i.e. (\ref{bi:01}) and input of only point A is used).
Both bi-directional and uni-directional predictions show good agreement with the experimental result until $t=100$.
However, the prediction accuracy of the uni-directional prediction becomes worse after this time because regressive waves reach measuring points.
On the other hand, the bi-directional prediction keeps its accuracy even after this time.
As the significant frequency of these irregular waves is $\omega_{1/3}=1.1$, prediction accuracy around this frequency is imperfect as shown in Fig. \ref{fig3}.
Nevertheless, the gap between prediction and experiment is acceptable.
Looking at Fig. \ref{fig3} (c) and (d), we can identify the time when regressive waves reach measuring points.
This information is helpful to conduct experiments on wave-related problems in the limited length of the water tank.
Note that it is not possible to validate the accuracy of predictions of progressive and regressive wave components because experimental results are not accessible.

\section{Conclusion}
As the prediction of incident waves is essential for controlling wave energy converters, deterministic decomposition and prediction methods of bi-directional water waves are studied.
Our wave prediction is based on the impulse response function.
We consider an array of three wave-rider buoys (A, P, and B) which are assumed to move in contact with the free surface of the water.
The use of the array of wave-rider buoys is in anticipation of the control of an array of point absorbers.
Using buoys on both sides (A and B), bi-directional waves are decomposed into progressive and regressive wave components.
The surface elevation of the middle buoy (P) is predicted by these decomposed progressive and regressive wave components.

This paper contains three new proposals.
(i) A new form of the impulse response function is proposed using a cosine-type window function to reduce an error due to the truncation of the infinite length of the function.
(ii) Decomposition and prediction methods of bi-directional water waves are formulated based on the discretization of the convolution integral, and the state-space representation of these methods are also given.
(iii) Generation method of initial uniform waves is shown.

To validate our proposals, numerical and experimental comparisons are demonstrated.
Both bi-directional regular and irregular waves are well decomposed into progressive and regressive wave components, and resultant wave predictions are also in good agreement.
As the impulse response function method can accurately predict wavefront (or transient waves), this may be superior to Fourier coefficients estimation in terms of time-series control.
This method can be also used for the tank experiment to detect reflected waves.

\section*{Acknowledgement }
This work was supported by JSPS KAKENHI Grant Number JP19K15218.



\bibliographystyle{elsarticle-num} 

\bibliography{main.bib}

\end{document}